# Magnetic-coupled electronic landscape in bilayer-distorted titanium-based kagome metals

Yong Hu[1,#,*], Congcong Le[2,#], Long Chen[3,4,#], Hanbin Deng[5], Ying Zhou[3,4], Nicholas C. Plumb[1], Milan Radovic[1], Ronny Thomale[6,7], Andreas P. Schnyder[8], Jia-Xin Yin[5], Gang Wang[3], Xianxin Wu[9,*], and Ming Shi[1,10,*]

[1] *Photon Science Division, Paul Scherrer Institut, CH-5232 Villigen PSI, Switzerland*
[2] *RIKEN Interdisciplinary Theoretical and Mathematical Sciences (iTHEMS), Wako, Saitama 351-0198, Japan*
[3] *Beijing National Center for Condensed Matter Physics and Institute of Physics, Chinese Academy of Sciences, Beijing 100190, PR China*
[4] *School of Physical Sciences, University of Chinese Academy of Sciences, Beijing 100190, PR China*
[5] *Department of physics, Southern University of Science and Technology, Shenzhen, Guangdong 518055, China*
[6] *Institute for Theoretical Physics, University of Würzburg, Am Hubland, D-97074 Würzburg, Germany*
[7] *Department of Physics and Quantum Centers in Diamond and Emerging Materials (QuCenDiEM) group, Indian Institute of Technology Madras, Chennai 600036, India*
[8] *Max-Planck-Institut für Festkörperforschung, Heisenbergstrasse 1, D-70569 Stuttgart, Germany*
[9] *CAS Key Laboratory of Theoretical Physics, Institute of Theoretical Physics, Chinese Academy of Sciences, Beijing 100190, China*
[10] *Center for Correlated Matter and Department of Physics, Zhejiang University, Hangzhou 310058, China*

[#] These authors contributed equally to this work.

[*] To whom correspondence should be addressed:

Y.H. (yonghphysics@gmail.com);

X.W. (xxwu@itp.ac.cn);

M.S. (ming.shi@psi.ch).

**Quantum materials whose atoms are arranged on a lattice of corner-sharing triangles, *i.e.*, the kagome lattice, have recently emerged as a captivating platform for investigating exotic correlated and topological electronic phenomena. Here, we combine ultra-low temperature angle-resolved photoemission spectroscopy (ARPES) with scanning tunneling microscopy and density functional theory calculations to reveal the fascinating electronic structure of the bilayer-distorted kagome material *Ln*Ti$_3$Bi$_4$, where *Ln* stands for Nd and Yb. Distinct from other kagome materials, *Ln*Ti$_3$Bi$_4$ exhibits two-fold, rather than six-fold, symmetries, stemming from the distorted kagome lattice, which leads to a unique electronic structure. Combining experiment and theory we map out the electronic structure and discover double flat bands as well as multiple van Hove singularities (VHSs), with one VHS exhibiting higher-order characteristics near the Fermi level. Notably, in the magnetic version NdTi$_3$Bi$_4$, the ultra-low base temperature ARPES measurements unveil an unconventional band splitting in the band dispersions which is induced by the ferromagnetic ordering. These findings reveal the potential of bilayer-distorted kagome metals *Ln*Ti$_3$Bi$_4$ as a promising platform for exploring novel emergent phases of matter at the intersection of strong correlation and magnetism.**

In recent years, there has been a burgeoning interest in two-dimensional (2D) quantum materials, where the 2D confinement of the electrons leads to enhanced many-body effects and novel symmetry-breaking states [1-8]. In these 2D materials a diverse set of intriguing quantum states has been observed, ranging from magnetism, charge density wave orders (CDW), unconventional superconductivity to electronic nematicity and beyond [1-8]. Among these 2D materials, compound composed of atoms arranged as a lattice of corner-sharing triangles, known as the kagome lattice, have emerged as a particularly captivating platform for exploring exotic correlated and topological electronic states [2-3,6-17].

Due to the intrinsic geometric frustration, the idealized 2D kagome network possesses intriguing electronic characteristics, such as flat bands, Dirac fermions, and van Hove singularities (VHSs). Recent experimental endeavors have primarily focused on transition-metal-based kagome magnets, exemplified by $Fe_mSn_n$ (m:n = 3:2, 1:1), $Co_3Sn_2S_2$, and $RMn_6Sn_6$ (where $R$ denotes a rare-earth element) [6,9-17], wherein the kagome layers are composed of magnetic atoms. Notably, in addition to the expected flat bands and Dirac fermions inherent to the hexagonal kagome structure [6,13,15], these materials have also been found to exhibit nontrivial Chern gaps [12] and relativistic Weyl fermions [9-11]. The interplay between such topological quantum states and intrinsic magnetism gives rise to a large anomalous Hall effect [9], underscoring the intriguing potential of these kagome magnets. Moreover, a recently discovered class of nonmagnetic kagome superconductors, such as $AV_3Sb_5$ (where $A$ can be K, Rb, or Cs), hosts a diverse nature of VHSs in the vicinity of the Fermi level ($E_F$) [18,19]. The large density of states (DOS) near these VHSs enhances correlations, which may lead to an intricate intertwining of CDW order and electronic nematicity [18-22]. Furthermore, by replacing vanadium with a kagome network of titanium atoms, a novel family of kagome superconductors, $ATi_3Bi_5$, was found to manifest electronic nematicity even in the absence of CDW order [23-25]. Nonmagnetic kagome lattices (whether vanadium- or titanium-based) present exciting prospects for exploring correlated quantum phenomena. Their derivatives, such as $RV_6Sn_6$ [26-28] and $LnM_3X_4$ (where $Ln$ represents Lanthanide; $M$ involves V, Ti; $X$ denotes Sb, Bi) [29-31], exhibit intriguing potential by introducing an anisotropic distortion to the kagome lattice and incorporating magnetic layers, thereby offering enhanced versatility and tunability and opening up intriguing avenues for exploration.

$LnTi_3Bi_4$ (where $Ln$ can be Nd, Yb, Pr, *etc*) is a newly synthesized titanium-based bilayer kagome metal [29-31], distinguished by its slightly distorted kagome lattices and zig-zag $Ln$ chains in $Ln$Bi bilayers. Despite the lack of long-range magnetic order in the kagome layer, akin to its analogs $ATi_3Bi_5$, the ability to introduce magnetism via controlling the $Ln$ sites offers a flexible platform for exploring inherent kagome lattice characteristics such as correlation effects, nontrivial band topology, and VHSs in combination with magnetism. Given the unique characteristics of distorted kagome lattices and tunable magnetism in zig-zag sublattices, the exploration of the electronic structure of such titanium-based kagome metal is desirable to showcase a prototypical electronic manifestation of novel Bloch eigenstates of the kagome lattice and to elucidate how magnetism manipulates the band topology, and as such to establish a new paradigm in the vibrant domain of kagome quantum materials.

In this work, we combine ultra-low temperature angle-resolved photoemission spectroscopy (ARPES), scanning tunneling microscopy (STM), and density functional theory (DFT) calculations to comprehensively examine the titanium-based bilayer kagome magnet NdTi$_3$Bi$_4$ alongside its nonmagnetic counterpart, YbTi$_3$Bi$_4$. Setting it apart from other kagome materials, *Ln*Ti$_3$Bi$_4$ exhibits a unique electronic structure characterized by two-fold symmetries ($C_{2x}$ and $C_{2z}$) originating from the distorted kagome lattice. We reveal the presence of double flat bands, and multiple VHSs, with one VHS very close to the $E_F$ displaying higher-order characteristics. Additionally, the ultra-low base temperature of the ARPES measurement enables us to observe an unusual band splitting of the band dispersion in the kagome magnet NdTi$_3$Bi$_4$ below its ferromagnetic (FM) ordering temperature. These findings underscore the potential of bilayer distorted kagome metals *Ln*Ti$_3$Bi$_4$ as a promising platform for exploring new emergent phases of matter at the intersection of strong correlation and magnetism.

*Ln*Ti$_3$Bi$_4$ possesses a layered crystal structure following the orthogonal space group *F mmm* (No. 69). It consists of Ti$_3$Bi$_4$ kagome layers alternating with *Ln*Bi$_2$ layers along the *c*-axis [Figs. 1a(i) and (ii)]. In contrast to the hexagonal structures of *A*V$_3$Sb$_5$ or *A*Ti$_3$Bi$_5$ [Fig. 1b(i)], *Ln*Ti$_3$Bi$_4$ features bilayer kagome layers within its primitive cell. Notably, these kagome layers display an additional anisotropic distortion, resulting in a transformation of the crystal symmetry from hexagonal to orthorhombic. Consequently, in our analysis we adopt the three-dimensional orthorhombic bulk Brillouin zone (BZ) [Fig. 1a(iii)], while the corresponding projected 2D BZ can be considered as being a pseudo-hexagonal, see the *Γ*, *M*, *K*, *M'* and *K'* high-symmetry points in Fig. 1a (iii), 1b(ii) and (iii). The reduction of crystal symmetry derived from the lattice distortion can influence the electronic structure features of the kagome lattice, such as Dirac cones and flat bands. Considering a single-orbital tight-binding model without spin–orbit coupling, a pristine single-layer kagome lattice [Fig. 1b(i)] is expected to host a gapless Dirac cone (protected by crystal symmetry) and a flat band, as illustrated in Fig. 1c(i). However, when an anisotropic in-plane distortion is introduced in the single-layer kagome lattice, the six-fold rotational symmetry ($C_{6z}$) is broken while the two-fold rotational symmetry ($C_{2z}$) is preserved [Fig. 1b(ii)], leading to a gap at the Dirac cone and a slight dispersion in the flat band [Fig. 1c(ii)] [6]. In the scenario of a bilayer distorted kagome lattice [Fig. 1b(iii)], the interlayer coupling between the two kagome layers further enriches the electronic structure, giving rise to double flat bands arising from bonding–antibonding splitting [Fig. 1c(iii)].

To explore the lattice-driven electronic confinement and double flat bands, we focus on two key members within the *Ln*Ti$_3$Bi$_4$ family: NdTi$_3$Bi$_4$, an FM kagome metal with a critical transition temperature ($T_c$) of 8.5 K, and the nonmagnetic YbTi$_3$Bi$_4$ (see Supplementary Fig. S1 for sample details). We first investigate the electronic structure of the two compounds through DFT calculations.

In line with our earlier findings in the titanium-based single-layer kagome metal $A$Ti$_3$Bi$_5$ [25], the theoretical orbital-resolved band dispersions, for both NdTi$_3$Bi$_4$ (Fig. 1d) and YbTi$_3$Bi$_4$ (Fig. 1e), display multiple $d$-orbital characteristics near the $E_\text{F}$. However, a distinct departure from previously studied kagome compounds is apparent in the overall band dispersions of NdTi$_3$Bi$_4$ and YbTi$_3$Bi$_4$, which display a highly anisotropic nature. This anisotropy is evident when comparing two high-symmetry cuts taken along the Γ - $M$ - $K'$ - Γ and Γ - $M'$ - $K$ - Γ directions (Fig. 1d and e), with Γ - $M$ and Γ - $K$ lines serving as two-fold rotational axes. Remarkably, the DFT calculations not only validate the expected presence of double flat bands below $E_\text{F}$, as illustrated in Fig. 1b(iii), but also unveil multiple VHSs in the vicinity of $E_\text{F}$ at the $M$ and $M'$ points. A detailed analysis of orbital character and band parity (Fig. S2) pinpoints the $d_{xy}/d_{x^2-y^2}$ orbitals as the primary contributors to these double flat bands [Figs. 1d(i) and 1e(i)]. Furthermore, the DOS calculation reveals a notable peak within the binding energy ($E_\text{B}$) range of 0.4 eV to 0.8 eV, further confirming the assignment of the $d_{xy}/d_{x^2-y^2}$ bands to the discussed double flat bands (highlighted by the grey stripe in Fig. 1d and e). These theoretical insights align consistently with our experimental observations, as we will demonstrate in the following.

Figure 2a presents atomically resolved STM topography measured on the surface of NdTi$_3$Bi$_4$, clearly revealing a hexagonal lattice formed by the Nd atoms. We note that only the Nd-terminated layer remains robust and dominant on the cleaved sample surface. To gain insight into the electronic structure of the system, we utilize quasiparticle interference (QPI) imaging, which manifests as static modulations in dI/dV(r, V) maps. The Fourier transforms of these maps show an isotropic scattering pattern, displaying an apparent Mirror symmetry (Fig. 2b and Fig. S3). Further exploring the isotropic nature of the electronic structure in NdTi$_3$Bi$_4$, Fig. 2c(i) showcases the experimental Fermi surface (FS) measured by ARPES on the Nd-terminated surface. It's evident that the FS topology exhibits a two-fold symmetry, disregarding the spectral intensity due to its sensitivity to the matrix element effect. Remarkably, this experimental kagome fermiology aligns well with the DFT calculations depicted in Fig. 2c(ii). Moreover, the experimental and theoretical FSs for YbTi$_3$Bi$_4$ (Fig. 2d) show a similar two-fold symmetry observed in NdTi$_3$Bi$_4$. This anisotropy in the low-energy electronic structure is further corroborated by the corresponding ARPES spectra acquired along two representative high-symmetry directions: the $\overline{M}$ - $\overline{\Gamma}$ - $\overline{M}$ and $\overline{M}'$ - $\overline{\Gamma}$ - $\overline{M}'$ paths, which are related by a six-fold rotation. These consistent observations across different samples, using different techniques, strongly suggest that the observed anisotropy in the electronic structure is due to the distorted kagome lattice contained in these systems.

We then turn to the search for characteristic band features, focusing specifically on the double flat bands and VHSs. In Fig. 3a and c, we display the polarization-dependent band dispersions of NdTi$_3$Bi$_4$ and YbTi$_3$Bi$_4$ along the $\overline{\Gamma}$ - $\overline{K}'$ direction, respectively. The presence of the double flat bands is clearly visible [Fig. 3a(i) and c(i)], particularly pronounced when probed with linear-vertical polarized light [Fig. 3a(iii) and c(iii)]. Our calculations well reproduce these double-flat-band features (Fig. 3b and d).

Intriguingly, the experimental double flat bands in NdTi$_3$Bi$_4$ (Fig. 3a) exhibit more apparent dispersions as compared to those in YbTi$_3$Bi$_4$ (Fig. 3c). This disparity can be attributed to the stronger inter-orbital hopping embedded in the kagome sublattice of NdTi$_3$Bi$_4$. Nonetheless, the observation of double flat bands in NdTi$_3$Bi$_4$ and YbTi$_3$Bi$_4$ (Fig. 3a-d), in contrast to the single flat band observed in the single kagome network RbTi$_3$Bi$_5$ (Fig. 3e and f), signifies the unique kagome-driven electronic confinement and hints at a potentially stronger electronic correlation effect in the double kagome motif.

The correlation effect is also embodied in the VHSs derived from the kagome lattice. Figures 3g and h show the energy-momentum dispersions of NdTi$_3$Bi$_4$ along the $\bar{\Gamma}$ - $\bar{M}$ - $\bar{K}'$ - $\bar{\Gamma}$ (Fig. 3g) and $\bar{\Gamma}$ - $\bar{M}'$ - $\bar{K}$ - $\bar{\Gamma}$ (Fig. 3h) directions from photoemission experiment, identifying multiple VHSs at the BZ boundary ($\bar{M}$ and $\bar{M}'$). The saddle point nature of these VHSs is further visualized through a series of cuts taken vertically across the $\bar{K}'\bar{M}$ [#$M_1$ – #$M_5$, as indicated in the inset of Fig. 3g(i)] and $\bar{K}\bar{M}'$ paths [#$M'_1$ –#$M'_5$ indicated in Fig. 3h(i)], where the band dispersions are hole-like along the $\bar{M}\bar{\Gamma}/\bar{M}'\bar{\Gamma}$ direction but electron-like along the $\bar{M}\bar{K}'/\bar{M}'\bar{K}$ direction. The band tops of the hole-like bands (dashed curves in Fig. 3i and j) exhibit minimum energies at the $\bar{M}$ and $\bar{M}'$ points (solid curves). The experimental VHS bands are overall consistent with the theoretical calculations [Fig. 3g(ii) and h(ii)], including the distinct anisotropies in band dispersions along the two high-symmetry directions (Fig. 3g and h). However, we note that the energy locations of the VHSs are not accurately captured by DFT, even including the consideration of an overall renormalization. Additionally, one VHS in the experiment, located very close to $E_F$, appears to exhibit a much flatter dispersion than within our DFT calculations [Fig. 3k(i)]. Upon fitting the experimental band dispersion, we find that the quadratic term is much smaller than the quartic one along the $\bar{M}'\bar{\Gamma}$ direction, indicating a higher-order nature of the VHS [Fig. 3k(ii)] [18,19]. These findings suggest that an orbital-dependent correlation effect is at play in the kagome magnet NdTi$_3$Bi$_4$.

Finally, we investigate how magnetism modifies the electronic structure in NdTi$_3$Bi$_4$, through detailed temperature-dependent ARPES measurements spanning across the magnetic transition temperature $T_c$. As summarized in Fig. 4a and b, upon cooling, the band dispersions initially display minimal changes [Fig. 4a(i, ii) and b(i, ii)]. As the temperature continues to decrease, however, a noticeable splitting in the β band along the $\bar{\Gamma}$ - $\bar{K}'$ direction becomes evident [Fig. 4a(iii, iv) and b(iii, iv)]. Upon warming, this splitting vanishes [Fig. 4a(v) and b(v)], and the bands revert to their initial states [Fig. 4a(i) and b(i)]. A similar band splitting also occurs in the temperature evolution of the band structure along the $\bar{\Gamma}$ - $\bar{M}'$ direction (Fig. 4c and d). Intriguingly, the band splitting manifests in a different band, namely, the α band. Furthermore, a side-by-side comparison between the constant energy contour take at $E_B$ = 0.15 eV above and below $T_c$ clearly reveals highly anisotropic band splittings in the magnetic state (Fig. 4e). Specifically, the α band exhibits maximum splitting along the $\bar{\Gamma}$ - $\bar{M}$ direction and minimal splitting along the $\bar{\Gamma}$ - $\bar{K}$ direction, whereas the β band displays the largest splitting along the $\overline{\Gamma}$-$\overline{K}$ direction and less pronounced splitting along the $\overline{\Gamma}$-$\overline{M}$ direction. These contrasting band splittings imply distinct magnetic couplings within these two bands.

To understand the origin of the observed band splittings in the specific α and β bands at low temperature, we turn to theoretical modelling, starting with an exploration of the orbital-resolved band structure in the paramagnetic phase. As depicted in Fig. 4f, the β and α bands are found to be dominantly contributed by the Nd *d* orbitals and the *p* orbitals of Bi atoms bonded to Nd atoms, respectively (for details, see Fig. S4). These orbital characteristics, coupled with the coincidence of the splitting temperature and $T_c$, near 8.5 K, suggest that the observed orbital-selective splittings are intimately related to the formation of FM order in the NdBi$_2$ layer (Fig. 4g). To simulate the effects of FM order, we calculate the band structures of NdTi$_3$Bi$_4$ by including a FM splitting within the Nd *d* orbitals and the Bi *p* orbitals that are bonded to the Nd atoms. Remarkably, these calculations replicate well the orbital-selective splittings (Fig. 4h) even though the theoretical calculations do not entirely capture the pronounced anisotropic nature of the splittings. One possible explanation for this discrepancy is the enhanced impact of the spin-orbit coupling associated with the surface NdBi$_2$ layers, a factor that has not been taken into account in our calculations.

In summary, we have investigated the intricate electronic band structure of the distorted bilayer kagome material *Ln*Ti$_3$Bi$_4$, which is characterized by two-fold symmetries due to the lattice distortion. In particular, we have unveiled the presence of double flat bands and multiple VHSs one of which exhibits a higher-order nature in close proximity to $E_F$. Additionally, we have observed an unconventional band splitting in the band dispersions of the kagome magnet NdTi$_3$Bi$_4$ below its FM transition temperature. The discovery of the double kagome flat bands calls for further research to attempt raising them to the $E_F$, which would allow to look for the proposed lattice-borne correlated topological phases [32]. While the origin of the anisotropic band splitting warrants further investigation, our results point toward the potential for generating magnetism-driven emergent spin-splitting surface states in the context of kagome materials [33]. Collectively, our findings establish the bilayer-distorted kagome metals *Ln*Ti$_3$Bi$_4$ as a promising platform for studying the intricate interplay between lattice distortion, electronic correlations, and magnetism within the framework of a distorted kagome lattice.

## Methods

**Single crystal growth and *in situ* doping**

*Ln*Ti$_3$Bi$_4$ (*Ln* = Nd and Yb) single crystals were grown by a high-temperature solution method using Bi as flux reported elsewhere [29,30]. Considering the possible air sensitivity of the surface, all manipulations and specimen preparation for structure characterization and property measurements were handled in an argon-filled glove box. X-ray diffraction data were obtained using a PANalytical X'Pert PRO diffractometer (Cu Kα radiation, λ = 1.54178 Å) operated at 40 *kV* voltage and 40 *mA* current with a graphite monochromator in a reflection mode (2θ = 5°–100°, step size = 0.017°). Indexing and Rietveld refinement were performed using the DICVOL91 and FULLPROF programs. The morphology and analyses of elements were characterized using a scanning electron microscope (SEM-4800, Hitachi) equipped with an electron microprobe analyzer for semiquantitative elemental analysis in energy-dispersive spectroscopy (EDS) mode.

**ARPES measurements**

Angle-resolved photoemission (ARPES) measurements were carried out at the ULTRA endstation of the Surface/Interface Spectroscopy (SIS) beamline of the Swiss Light Source, using a Scienta Omicron DA30L analyzer. The measurements were performed with a total energy resolution of 15 *meV*. The samples were cleaved in-situ with a base pressure of better than $5 \times 10^{-11}$ *torr*, and measured at 20 *K*. The Fermi level was determined by measuring a polycrystalline Au in electrical contact with the samples.

**STM measurements**

The scanning tunneling microscopy (STM) results are obtained with a Unisoku USM-1300 system equipped with a low temperature cleaving stage. Single crystals are cleaved mechanically in situ at 77 *K* in ultra-high vacuum conditions (better than $2 \times 10^{-10}$ *torr*), and then immediately inserted into the STM head at 5 *K*. Topographic images are obtained with Ir/Pt tips with V = 1 *V*, I = 1 *nA*. The conductance maps are acquired by taking a spectrum at each location with feedback loop off. Standard lock-in method is used in the spectrum measurement with junction set up of V = -100 *mV*, I = 1 *nA* and oscillation voltage of 5 *mV* unless otherwise noted. The frequency of the oscillation voltage is 973.231 *Hz*. The data is mirror symmetrized if noted in the text.

**Computational Methods**

Our calculations are performed using density functional theory (DFT) as implemented in the Vienna *ab initio* simulation package (VASP) code [34-36]. The Perdew-Burke-Ernzerhof (PBE) exchange-correlation functional and the projector-augmented-wave (PAW) approach are used. Throughout this work, the cutoff energy is set to 500 *eV* for expanding the wave functions into plane wave basis, and spin-orbital coupling (SOC) is included. The Brillouin zone is sampled in the *k* space within the Monkhorst-Pack scheme [37], and the *k* mesh used is 8 × 8 × 8 on the basis of the equilibrium structure. To get the tight-binding Hamiltonian, we create Wannier functions via the package Wannier90 [38] with a projection of the Bloch states to the atomic orbitals, and then use this Hamiltonian to calculate surface states [39]. In our calculations, we adopt the experimental and optimized structural parameters for NdTi$_3$Bi$_4$ and YbTi$_3$Bi$_4$, respectively.

**Acknowledgement**

The work was supported by the Swiss National Science Foundation under Grant. No. 200021_188413. Y.H. acknowledges the support from the National Natural Science Foundation of China (Grant No. 12004363). X.W. was supported by the National Natural Science Foundation of China (Grant no. 12047503)

**Author Contributions Statement**

Y.H. conceived the ARPES experiments. L.C. grew and characterized the crystals with guidance from Y.Z. and G.W.. X.W. and C.L. performed the theoretical calculations with the support from R.T. and A.P.S.. Y.H. performed the ARPES experiments with support from N.C.P., M.R. and M.S.. H.D. and J.-X.Y. performed the STM experiments. N.C.P. maintained the ARPES facilities at the SIS-ULTRA. Y.H. analyzed the data. Y.H. wrote the paper with inputs from all authors. M.S., X.W. and Y.H. supervised the project.

**Competing Interests Statement**

The authors declare no competing interests.

**Data availability**

The data that support the findings of this study are available from the corresponding authors upon reasonable requests.

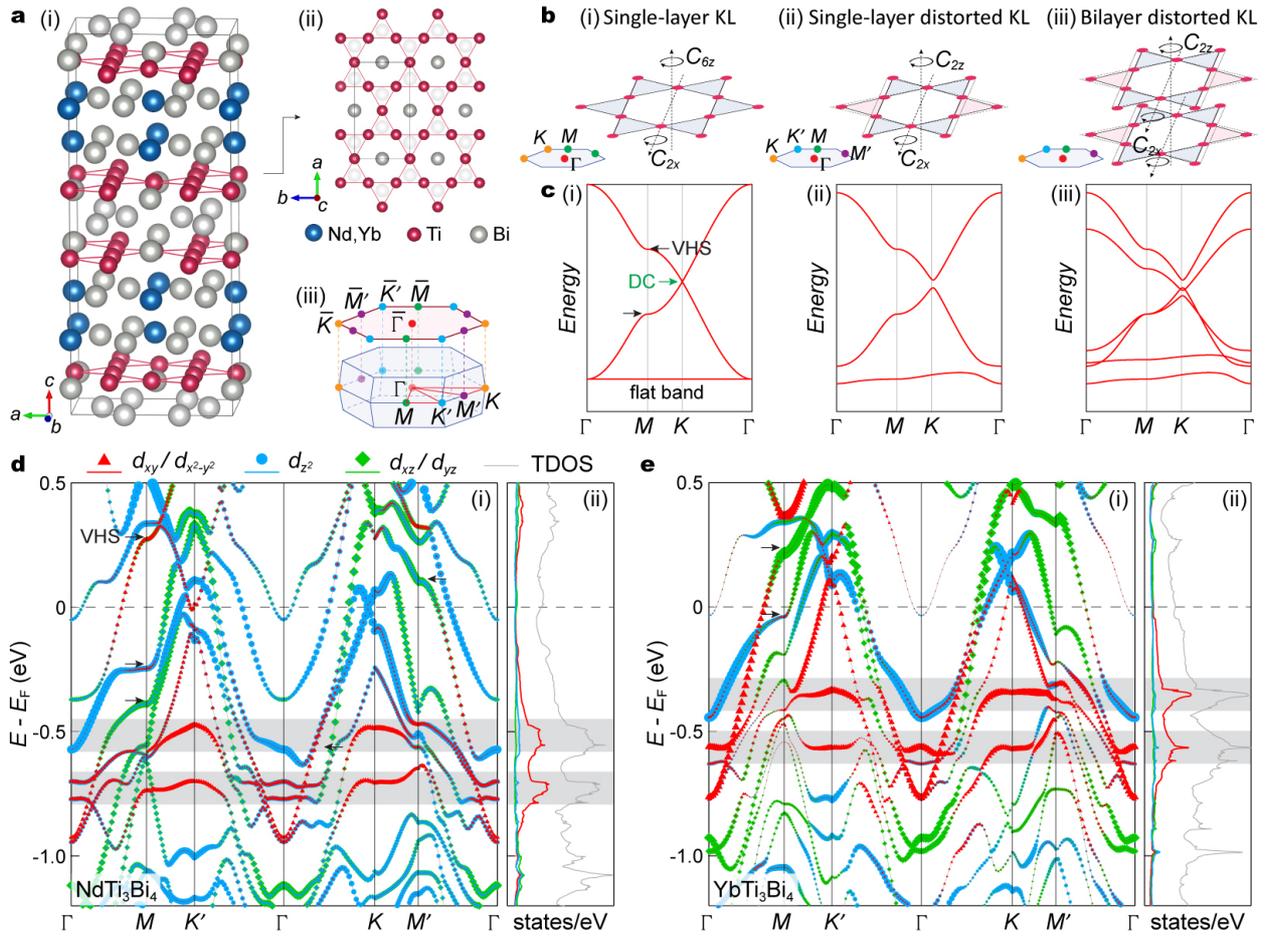

**Fig. 1 | Crystal structure and calculated band structure of $Ln$Ti$_3$Bi$_4$.** **a** The unit cell of $Ln$Ti$_3$Bi$_4$ (i), a top view displaying the kagome plane (ii), and the bulk BZ of $Ln$Ti$_3$Bi$_4$ with the corresponding projected BZ (iii). **b** Two-fold ($C_{2x}$) and six-fold ($C_{6z}$) rotation-axis symmetries of a pristine single kagome layer (KL) (i), two-fold symmetries ($C_{2x}$ and $C_{2z}$) of the single-layer distorted KL (ii), and bilayer distorted KL (iii). **c** A single-orbital tight-binding model without spin–orbit coupling for the single KL (i), single-layer distorted KL (ii), and bilayer distorted KL (iii). **d,e** DFT calculations representing orbital character-resolved band structures of NdTi$_3$Bi$_4$ [d(i)] and YbTi$_3$Bi$_4$ [e(i)] alongside their corresponding DOS calculations [d(ii), e(ii)].

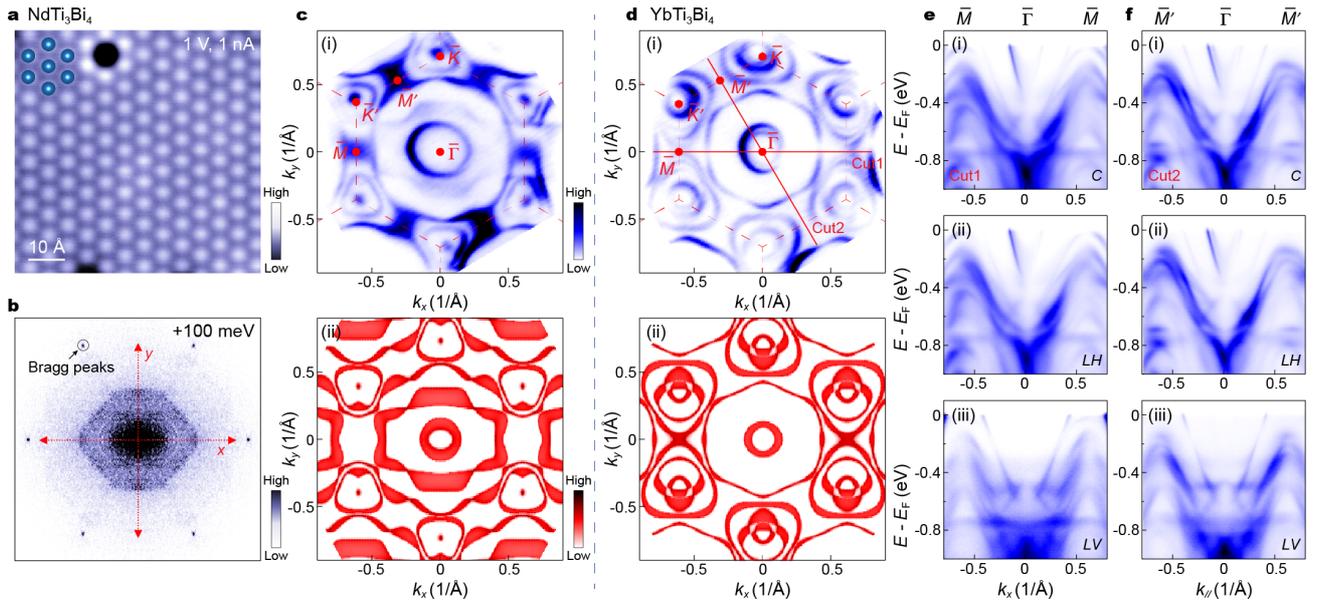

**Fig. 2 | Anisotrpic electronic structure in *Ln*Ti$_3$Bi$_4$. a,b** STM topograph of Nd termination measured at 5 *K* (a) and corresponding Fourier transforms (b). Atomic Bragg peaks are highlighted with a black ring. **c** Experimental constant energy contour (CEC) (i) and theoretical CEC (ii) of NdTi$_3$Bi$_4$. **d** Same data as in (c), but for YbTi$_3$Bi$_4$. **e,f** ARPES spectra along the $\overline{\Gamma}$-$\overline{M}$ (e) and $\overline{\Gamma}$-$\overline{M}'$ (f) directions [red lines in d(i)], probed with circularly (*C*) (i), linear horizontally (*LH*) (ii), and linear vertically (*LV*) (iii) polarized lights.

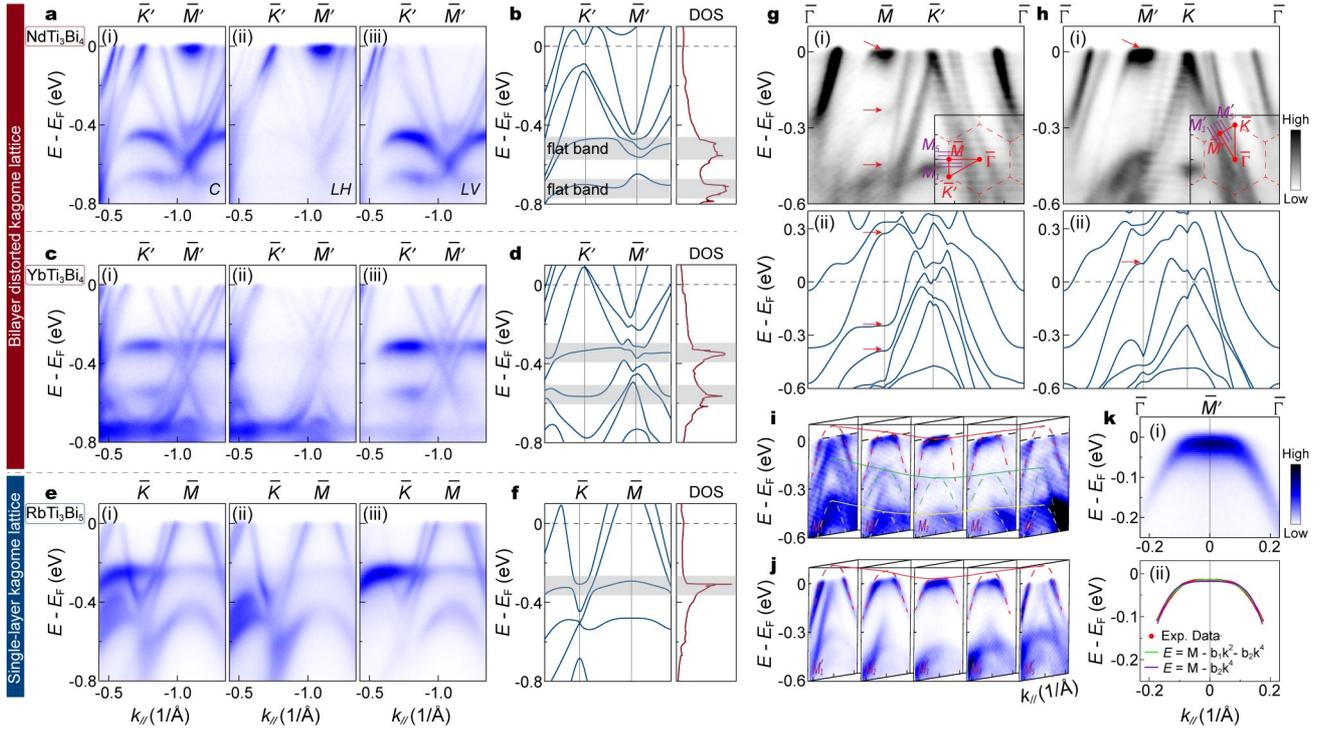

**Fig. 3 | Double flat bands and multiple VHSs in *Ln*Ti₃Bi₄. a** Experimental band dispersions of NdTi₃Bi₄ measured with *C* (i), *LH* (ii), and *LV* (iii) polarized lights along the $\overline{\Gamma}$- $\overline{K}'$ - $\overline{M}'$ direction. The grey shading highlights the flat band. **b** DFT-calculated band structure and DOS of NdTi₃Bi₄ along the $\overline{\Gamma}$- $\overline{K}'$ - $\overline{M}'$ direction. **c-f** Same data as in (a,b), but for YbTi₃Bi₄ (c,d) and RbTi₃Bi₅ (e,f). **g,h** Experimental (i) and theoretical (ii) band dispersions of NdTi₃Bi₄ taken along the $\overline{\Gamma}$ - $\overline{M}$ - $\overline{K}'$ - $\overline{\Gamma}$ (g) and $\overline{\Gamma}$ - $\overline{M}'$ - $\overline{K}$ - $\overline{\Gamma}$ (h) directions [red lines in Fig. 1a(iii)]. The arrows indicate the VHSs. **i,j** ARPES spectra taken vertically across the $\overline{M}$ - $\overline{K}'$ (i) and $\overline{M}'$ - $\overline{K}$ (j) paths, with the paths of the band dispersions (#M₁-#M₅) indicated by the purple lines in the insets of (g, h). **i** Experimental band dispersion of NdTi₃Bi₄ along the $\overline{\Gamma}$- $\overline{M}'$ direction (i), fittings of the measured dispersion by $E = M - b_1 k^2 - b_2 k^4$ and $E = M - b_2 k^4$ forms (ii).

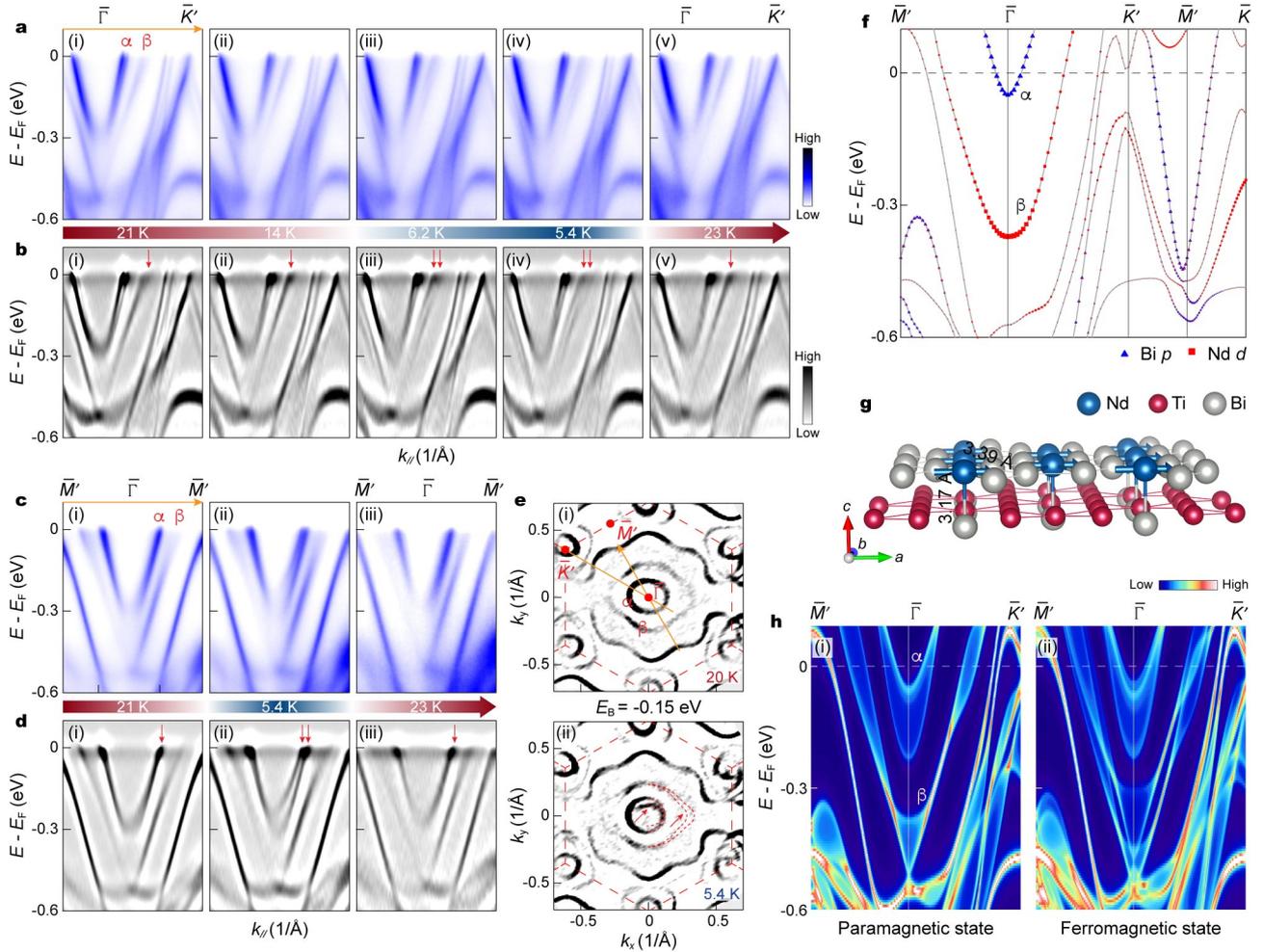

**Fig. 4 | Magnetism-driven orbital-selective band splittings in NdTi$_3$Bi$_4$. a,b** Temperature evolution of the band dispersions along the $\overline{\Gamma}$-$\overline{K}'$ direction, probed with *C* polarized light (a), and their corresponding second-derivative plots with respect to energy (b). **c,d** Same data as in (a,b), but measured along the $\overline{\Gamma}$-$\overline{M}'$ direction. **e** CEC taken at $E_B$ = 0.15 eV above (i) and below (ii) the magnetic transition temperature. The yellow lines mark the momentum path of the band dispersion shown in (a-d). The red dashed curves and arrows highlight the orbital-selective anisotropic band splittings. **f** Nd-*d* and Bi-*p* orbital-resolved DFT band structure. **g** Ferromagnetic NdBi$_2$ layer and its adjacent Ti$_3$Bi$_4$ kagome layer. **h** Calculated projection of bulk states in paramagnetic state (i) and FM state by introducing a FM order within the Nd *d* orbitals and the Bi *p* orbitals bonded to Nd atoms (ii).

14